\documentclass[final,5p,times,twocolumn]{elsarticle}
\usepackage{graphicx}
\usepackage{amssymb}
\usepackage{amsthm}
\usepackage{amsmath}
\usepackage{hyperref}
\usepackage{natbib}

\usepackage{xcolor}
\usepackage{amsmath}
\journal{Journal of Nuclear Materials}

\begin{document}

\begin{frontmatter}

\title{On the detection and classification of material defects 
in crystalline solids after energetic particle impact simulations}
%\tnotetext[mytitlenote]{Fully documented templates are available in the elsarticle package on \href{http://www.ctan.org/tex-archive/macros/latex/contrib/elsarticle}{CTAN}.}

%% Group authors per affiliation:
\author[ipp]{F. Javier Dom\'inguez-Guti\'errez \corref{cor1}}
\ead{javier.dominguez@ipp.mpg.de}
\cortext[cor1]{Corresponding author}
\author[ipp]{U. von Toussaint}
\address[ipp]{Max-Planck Institute for Plasma Physics, Boltzmannstrasse 2, 85748
 Garching, Germany.}

%% or include affiliations in footnotes:
%\author[mymainaddress,mysecondaryaddress]{Elsevier Inc}
%\ead[url]{www.elsevier.com}

\begin{abstract}
We present a fingerprint-like method to analyze material defects 
after energetic particle irradiation by computing a rotation 
invariant descriptor vector for each atom of a given sample. 
For ordered solids this new method is easy to use, does not require extreme computational 
resources, and is largely independent of the sample material and sample temperature. 
As illustration we applied the method to molecular dynamics simulations of deuterated 
and pristine tungsten lattices at 300 K using a primary knock-on atom (PKA) of 1 keV with different velocity 
directions to emulate a neutron bombardment process.
The number of W atoms, that are affected after the collision cascade,  
have been quantified with the presented approach.
At first atoms at regular lattice positions as well as common defect types like 
interstitials and vacancies have been identified using precomputed descriptor 
vectors. 
A principal component analysis (PCA) is used to identify previously overlooked 
defect types and to derive the corresponding local atomic structure.
A comparison of the irradiation effects for deuterated and pristine tungsten 
samples revealed that deuterated samples exhibit consistently more defects 
than pristine ones.

\end{abstract}


\end{frontmatter}

%\linenumbers

%% References with bibTeX database:
\section{Introduction}
\label{intro}

Plasma facing materials (PFMs) in the next generation of nuclear reactors are
subjected to extreme conditions due to the direct plasma interaction 
\cite{doi:10.13182/FST11-A12443}.
%In order to study optimal PFMs, Molecular dynamics (MD)-simulations provide a p
%owerful approach 
%to simulate radiation damage in selected materials.
The induced material damage has to be analyzed by the identification,
classification, and quantification of lattice defects. 
This is a complex analysis, complicated by potentially 
unforeseen atomic configurations after irradiation \cite{BOLT200243,mayer}.

Material defects on an atomic scale are often investigated using Voronoi tessellation 
and the analysis of Wigner-Seitz cell volumes to identify common defects like vacancies and interstitials  
\cite{Atsuyuki,doi:10.1063/1.4849775}. Also a 
description of the topology of individual grains, bubbles, and cells in 
three-dimensional polycrystals \cite{PhysRevLett.109.095505} is sometimes provided.
However, non-standard and unexpected material defects provide a challenge to the existing methods. A robust method is needed to classify 
and quantify damage e.g. after the
impact of high-energy particles from fusion reactions.
For this reason, the International Atomic Energy Agency set out a 
competition \cite{IAEA} to detect material defects in different metals. 
In this paper, we present a novel method to identify, classify, and 
quantify material defects regardless of the sample composition and temperature, 
and impact energy. 
Our numerical procedure is able to identify new material defect types other 
than the common ones like interstitial sites and vacancies. 
It is based on the Smooth Overlap of Atomic positions 
method \cite{PhysRevB.87.184115,PhysRevB.90.104108} (SOAP) to compute the 
descriptor vector of individual atoms of a given sample, which 
describes their particular physical environment. 
SOAP is implemented in the QUantum mechanics and Interatomic 
Potentials package (QUIP) \cite{quip} that is a collection of software tools 
with a variety of interatomic potentials and tight binding packages.
As an example, we analyze the defects in single-crystal tungsten after the 
interaction with a 1 keV of primary knock-on atom, due to tungsten is an important
candidate as PFM for a fusion device \cite{KAUFMANN2007521}.
It is also of interest to investigate the effect of deuterium
decoration of W on the material defects formation
\cite{PIAGGI2015233,CHEN2016190}.

This paper is organized as follows: In section \ref{methods}, we 
present the computational methods that are utilized in our research work, 
followed by a description of the preparation of the numerical cells used 
to perform MD simulations of a deuterated and pristine W lattice for 
1 keV of primary knock-on atoms.
We also briefly describe the computation of the descriptor vectors that define
our crystal defects classifications, quantification and further results for the
different systems are shown in the section \ref{results}. 
Finally, in section \ref{conclusions}, we discuss our results and 
provide concluding remarks.

%%%%%%%%%%%%%%%%%%%%%%%%%%%%%%%%
\section{Theory}
\label{methods}

\subsection{Descriptor vectors}
An ideal tool for the identification of distortions in a crystalline sample 
should be sensitive to changes in a local environment, but at the same time 
insensitive to "global" changes like rotation of the sample or permutations 
of its atoms. 
In addition, it should allow for some kind of calibration, e.g. not 
counting thermal fluctuations as defects. Furthermore, the possibility of
 a probabilistic interpretation instead of a binary yes/no-response would 
be beneficial. Most of the suggested approaches so far fail in one or more 
of these aspects or can only be applied to specific kinds of defects.
For example, just using the Cartesian coordinates of the neighboring atoms 
would result in a false positive identification of a changed environment 
under simple rotation. 
Similarly the use of a vector of nearest-neighbor distances (which are 
invariant under rotation) has deficiencies, where very different atomic 
configurations can result in the same vector 
\cite{PhysRevB.87.184115}.

In this section, we outline the approach to compute the 
reference descriptor vector based on the QUantum mechanics and Inter-atomic 
Potentials package (QUIP) \cite{quip}, that can be considered as a footprint 
of the local environment of an atom \cite{PhysRevB.87.184115,PhysRevB.90.104108}. 

\subsection{Descriptor vector computation}
The analysis of the defects of a given material structure is 
started by first computing the descriptor vectors (DVs) of all the atoms 
in the sample.
The DV of the $i$-th atom of the sample, $ \vec{\xi}^{\ i}$ (defined below), 
is a component vector and a representation of the atomic neighborhood within 
some cutoff-radius. 
The representation is based on an expansion of the local neighbourhood of an
atom in a product of spherical harmonic functions and radial basis functions. 
It has been shown in Ref. \cite{PhysRevB.87.184115} that this representation
exhibits the desired properties, i.e. that the descriptor vector introduced 
below is invariant to rotation, reflection, translation, and permutation of 
atoms of the same species, but sensitive to small changes in the local atomic
environment \cite{PhysRevB.87.184115}.
%.which retains the Cartesian 
%representation of each atom with a highly dependence of the atoms 
%location in the sample.
%Besides, we overcome the angular wave number dependence by computing 
%the DVs with Smooth Overlap of Atomic Positions (SOAP) method 
%\cite{PhysRevB.87.184115,PhysRevB.90.104108}, 
In order to computed the DVs, let us define the 
atomic environment of an $i$-th atom by a sum of truncated Gaussian density 
functions as  \cite{PhysRevB.87.184115},

\begin{equation}
\rho^{{} i}(\vec r) = \sum^{\textrm{neigh.}}_{j} \exp 
\left( -\frac{|\vec r-\vec r^{\ ij}|^2}{2 \sigma^2_{\textrm{atom}}} \right) 
f_{\textrm{cut}} \left( |\vec r^{\ ij}| \right),
\label{eq:Eq1}
\end{equation}
%where the $\vec r$ is the position vector of the $i$-atom of the sample, 
%and $\vec r_{ij}$ is the neighboring atomic position vectors.
where $\vec r^{\ ij}$ denotes 
the difference vector between the atom positions $i$ and $j$.
$\sigma^2_{\textrm{atom}}$ denotes the broadening of the atomic position, 
this parameter is set according to the 
lattice constant of the sample and takes into account the thermal motion.
Finally, $f_{\textrm{cut}} \left( |\vec r^{\ ij}| \right)$ is a smooth 
cutoff function, which is required to limit the considered neighborhood 
of an atom. 
%within a finite cutoff distance due to the crystal structure.
Then, $\rho^{\ i}(\vec r)$ can then be expressed in a basis set, 
as obtained in \cite{PhysRevB.90.104108}, in terms of expansion coefficients, 
$c_{nlm}$: 
%This allows to compute the normalized descriptor vector $\hat q$ as
\begin{equation}
%\rho^i(\vec r) = \sum_{nlm}^{NLM} c^{(i)}_{nlm}g_n(r)Y_{lm}\left(\frac{\vec{r}}{\left|\vec{r}\ %\right|}\right),
\rho^i(\vec r) = \sum_{nlm}^{NLM} c^{(i)}_{nlm}g_n(r)Y_{lm}\left(\hat r\right),
\label{eq:Eq2}
\end{equation}
%\begin{equation}
%    \hat{r}=\frac{\vec{r}}{\left|\vec{r}\right|}.\nonumber
%\end{equation}
with $\hat r = \vec{r}/ \left|\vec{r}\right|$ as unit-vector in the direction 
of $\vec{r}$. 
The $c^{(i)}_{nlm} = \langle g_n Y_{lm} | 
\rho^i \rangle$ are the expansion coefficients that corresponds
to the $i$th-atom in the lattice and $g_n(r)$ is a set of orthonormal
radial basis functions $(\langle g_{n}(r)\mid g_{m}(r)\rangle = \delta_{nm} )$, and $Y_{lm}(\hat r)$ are the spherical harmonics  with the atom positions projected 
onto the unit-sphere. The inner product $\langle f\mid g\rangle$ used above is 
given by the integral over the surface of a unit-sphere
\begin{equation}
    \langle f \mid g\rangle = \int f^{*}\left(\hat{r}\right)g\left(\hat{r}\right)\mathrm{d}\Omega\left(\hat{r}\right).
\end{equation}

To achieve invariance against rotations of the local environment of atom $i$, 
Eq. \ref{eq:Eq2} needs to be averaged over all possible rotations. 
After some algebra \cite{PhysRevB.90.104108}, 
the desired rotation invariant result can be expressed by the 
multiplication of the $c_{n'lm}$ with the complex conjugate coefficient 
$c^{*}_{nlm}$, summed over all $m$.
This resembles some similarity to the power spectrum of Fourier coefficients.
Thus the components of the DV of the $i$-th atom, $ \vec{\xi}^{\ i}$, 
are given by  \cite{PhysRevB.90.104108}
\begin{equation}
\vec{\xi}^{\ i} = \left\{ \sum_m \left(c_{nlm}^i \right)^* c_{n'lm}^i \right\}_{\ n,n',l},
\label{eq:Eq3}
\end{equation}
where each component of the vector corresponds to one of the index 
triplets $\{n,n',l\}$. 
In the following we will use the normalized vector $\vec{q}^{\ i}$ as descriptor 
vector (DV) for the local environment of atom $i$:
$\vec{q}^{\ i} = \vec{\xi}^{\ i}/|\vec{\xi}^{\ i}|$.
%The SOAP kernel takes the value of unity only when the two neighborhoods 
%are identical.
We compute the DVs of the deuterated and pristine sample using the QUIP 
package with its python interface \cite{quip}. 

%%%%%%%%%%%%%%%%%%%%%%%%%%%%%%%%%%%%%%%%%%%%%%%%%%%%%%%%%%%%%%%%%%%%%%%%%%%%
%%%%%%%%%%%%%%%%%%%%%%%%%%%%%%%%%%%%%%%%%%%%%%%%%%%%%%%%%%%%%%%%%%%%%%%%%%%%
\subsection{Calibration of the descriptor vectors}
Using the sequence of Eq. \ref{eq:Eq1} to \ref{eq:Eq3} for each atom $i$ of 
the sample a DV, $\vec{q}^{\ i}$, can be computed. 
Depending on the choice of the expansion orders in Eq. \ref{eq:Eq2} for the 
spherical harmonics and the radial basis functions the number of components 
of $\vec{q}^{\ i}$ varies. 
Here, we used $n = 4$, and $l = 4$ (with $-l \leq m \leq l$) which yields a 
DV with $k=51$ $\left( 0 \dots 50  \right)$ components. 
\footnote{Besides the zeroth-component there are for each $l \in \{0,4\}$
10 ordered pairs (n,n'), i.e. $\left\{(1,1),(1,2),(1,3),(1,4),(2,2),(2,3),(2,4),(3,3),(3,4),(4,4)\right\}$, 
yielding a DV with $50+1$ components.}
%\textcolor{blue}{ Fig. (\ref{fig:fig_2})}.
The difference of two local environments of atom $i$ and atom $j$ can then be 
obtained by computing the distance $d$ between the two DVs, 
$d = d \left( \vec{q}^i, \vec{q}^j \right)$, which for the standard euclidean 
measure is defined as
$d^E = \sqrt{\sum_k \left( q^i_k - q^j_k \right)^2}$, where $k$ is used as 
a component index. However, giving all components the same weight may not always
be appropriate because some of them may be more fluctuating than others - 
although the standard euclidean distance works quite well in most cases that 
we have looked at.
%identical or a weighted measure  
%$d^M = \sqrt{\sum_k \left( q^{\ i}_k - q^{\ j}_k \right)^T \sum^{-1} \left( q^{\ i}_k - q^{\ j}_k \right)}$ i.e. the Mahalanobis distance
% measure \cite{Maha}.
In order to select an appropriate measure two compare the DVs, we used a 
MD simulation to generate a thermalized tungsten bcc lattice at $T = 300$ K 
(see section 3) without defects, and computed the DV for all the atoms. 
This group of DVs for a defect free and thermalized environment has been used 
to compute a mean reference descriptor vector as: $\vec{v}\left(T\right)=\frac{1}{N}\sum_{i=1}^{N}\vec{q}^{i}\left(T\right)$, as 
well as the associated
%the variance,$\sigma(T)$, and
covariance matrix $\Sigma\left(T\right)$ , where especially the covariance 
matrix depends on the sample temperature.
Following this, the distance difference of the DVs of the thermalized
environment from the mean bcc-lattice can be 
computed using the Mahalanobis distance \cite{Maha}
\begin{equation}
    d^M (T) \left(\vec{q}^{i},\vec{v}\left(T\right)\mid \Sigma \right)= \sqrt{ \left( \vec{q}^{\ i} - \vec{v}\left(T\right) \right)^T 
\Sigma^{-1} (T) \left( \vec{q}^{\ i} - \vec{v}\left(T\right) \right)}
\label{eq:maha}
\end{equation}
This yields to a distance distribution of the DVs for a thermalized and 
defect-free lattice, which sets out the scale to judge if an unexpected large 
distortion of the local environments is present. 
For the computation of the reference DVs for other common types of point defects 
a similar approach is chosen: A small numerical cell containing the defect of 
interest (e.g. an interstitial) was prepared and the computation of the 
DVs of the atoms defines the fingerprint for this specific atomic 
environment (see section 3).

%%%%%%%%%%%%%%%%%%%%%%%%%%%%%%%%%%%%%%%%%%%%%%%%%%%%%%%%%%%%%%%%%%%%%%%%%%%%%%
%%%%%%%%%%%%%%%%%%%%%%%%%%%%%%%%%%%%%%%%%%%%%%%%%%%%%%%%%%%%%%%%%%%%%%%%%%%%%%%%
\subsection{Probability of being a lattice atom }
The use of the Mahalanobis distance $d^M$(T) allows a
straightforward probabilistic interpretation of the distance. 
Given that the distance distribution  $d^M$(T) of a 
thermalized sample is close to a Gaussian distribution, then the probability,
$P\left(\vec q^{\ i} \mid \vec{v}\left(T\right)\right)$, 
of an atom $i$ being in a locally undistorted lattice can be computed using
%We can assign a probability, $P(\vec q^{A} | \langle y \rangle)$, of 
%finding a W atom in a body centered cubic (BCC) position as
\begin{equation}
\begin{split}
\centering
P\left(\vec q^{\ i} \mid \vec{v}\left(T\right)\right) & = 
P_0 \exp \left[ -\frac{1}{2}d^M(T)^2 \right],\\
d^M (T)^2 &=   \left( \vec q^{\ i} - \vec{v}\left(T\right)
\right)^T  \Sigma^{\Large{-1}}\left(T\right) 
\left( \vec q^{\ i} - \vec{v}\left(T\right) \right),
\label{eq:Eq4}
\end{split}
\end{equation}
where $P_0$ is the normalization factor and $ \vec{q}^{\ i}$
is the DV of atom $i$.  
For our present analysis we only use the diagonal elements of the covariance matrix, i.e.
set $\Sigma_{ij}=0\, \mathrm{for}\; i\neq j$.
%diagonal with the variances $\sigma^2$ as diagonal elements. 
This yields a quantitative measure for identifying/selecting defects, 
even for samples at different temperatures, where fixed criteria 
(e.g. maximum displacement) could easily fail.
%Here, we can take into account thermal motion by computing the DV 
%of each atom, $\vec y_i$, of a thermalized BCC sample to define a 
%mean DV, $\langle \vec y \rangle$, and co-variance matrix, 
%$\left( A_{i,i} \right)^{-1}$.
%On this way, we guarantee that a W atom is likely to have a lattice 
%position in the sample. 

%%%%%%
\subsection{MD simulations}
\begin{figure}[!t]
\centering\includegraphics[width=0.48\textwidth]{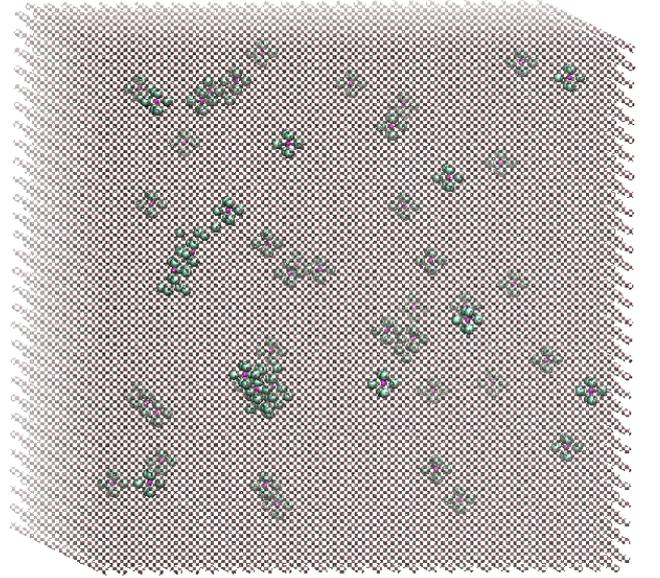}
\caption{(Color on-line) Deuterated W lattice. This 
simulation box has been energy optimized and thermalized to $300$ K.
Colors: W atoms in a bcc position are presented as light gray spheres; 
W atoms around a D atom as depicted as green spheres; and D atoms are 
illustrated as purple small spheres.}
\label{fig:fig_1}
\end{figure}

The simulation box is initially prepared as a pristine single-crystalline 
W lattice sample with 48778 W atoms based on a bcc unit cell.
The box has a dimension of $(28a,28a,28a)$ with $a = 0.316$ nm as the 
W lattice constant \cite{SETYAWAN2015329}, for a sample temperature 
of 300K due to the majority of the experiments of tungsten damaging are 
done at room temperature 
\cite{Wright_nucl_fusion,Herrmann_nucl_fusion}.
The deuterated W lattice is created by introducing $0.1$ at \% D (53 
deuterium atoms) randomly distributed in the whole sample at tetrahedral 
interstitial positions, Fig. \ref{fig:fig_1}.
All the samples are first energy optimized and subsequently thermalized to 
300 K, using a Langevin thermostat with a time constant of 100 fs
\cite{PhysRevB.17.1302,DOMINGUEZGUTIERREZ201756}.
%In Fig. \ref{fig:fig_1}, we show the prepared deuterated W lattice
%using the following color code: D atoms are represented as purple spheres; 
%W atoms that are next to a D atom are displayed as green spheres; and 
%lattice W atoms are shown as light-gray spheres to better visualize the 
%deuterated sample. 
%It is worth noting that the W atoms around the D atom are identified
%by our method.

%%%%%%%%%%%%%%%%%%%%%%%%%%%%%%%
%%%%%%%%%%%%%%%%%%%%%%%%%%%%%%%
MD simulations are performed by assigning an impact energy of 1 keV to a W atom,
thus acting as a primary-knock on atom (PKA). 
Its initial position is located at the center of the numerical box.
We consider ten velocity directions: $\langle 
0 0 1 \rangle$, $\langle 1 0 1 \rangle$, $\langle 1 1 1 \rangle$, and 7 
cases for $\langle r_1 r_2 r_3 \rangle$ where $r_i$ are random numbers 
uniformly distributed in an interval of $[0,1]$.
A velocity-Verlet integration algorithm is applied to model the collision
dynamics, which lasts for 10 ps with a time step of $\Delta t = 
1\times10^{-3}$ ps.
%This simulation time assures that the W sample will reach its
%thermal stability at the final step to find actual material defects.
After that time the dynamics of the collision cascade has stopped 
and subsequent diffusive relaxation process are thus missed, but their 
proper consideration would in any case exceed the time scale accessible 
for MD method.
The simulations were performed on a desktop computer using the 
Large-scale Atomic/Molecular Massively Parallel Simulator (LAMMPS code) 
\cite{PLIMPTON19951} with the reactive inter-atomic potential for 
the ternary system W-C-H of Juslin et al. \cite{Juslin}, which is
 based on an analytical bond-order scheme.
This potential has been used to study 
neutron damage in poly-crystalline tungsten \cite{1402-4896-2011-T145-014036},
trapping and dissociation processes of H in tungsten vacancies 
\cite{FU2018278}, and cumulative bombardment of low energetic H atom 
of W samples for several crystal orientations \cite{FU2018}.

%%%%%%
\section{Results}
\label{results}

%%%%%%
%\subsection{Main reference vectors}
%\label{reference_vectors}

%A perfect body-centered-cubic (BCC) unit cell, an atom in an tetrahedral 
%interstitial site, and an atom next to a void/vacancy are considered to 
%calculate the main reference descriptor vectors. 
Initially, we computed reference DVs for the thermalized defect free bcc lattices 
for two temperatures, and two of the most common defects, interstitial
atoms and W atoms next to single vacancy. 
For that, in order to compute the main reference vectors, we used a sample of 
432 W atoms in a  fully periodic box with lateral 
dimensions of (5$a$, 5$a$, 5$a$), based on a bcc unit cell of dimension $a = 0.316$ nm.
A tungsten atom is removed from this sample to calculate the 
reference vector for atoms next to a single vacancy.
For the interstitial case, a W atom is introduced into the lattice 
at a tetrahedral interstitial site. 

\begin{figure}[!b]
%\centering\includegraphics[width=250pt,height=150pt]{Figures/Fig2.eps}
\centering\includegraphics[width=0.48\textwidth]{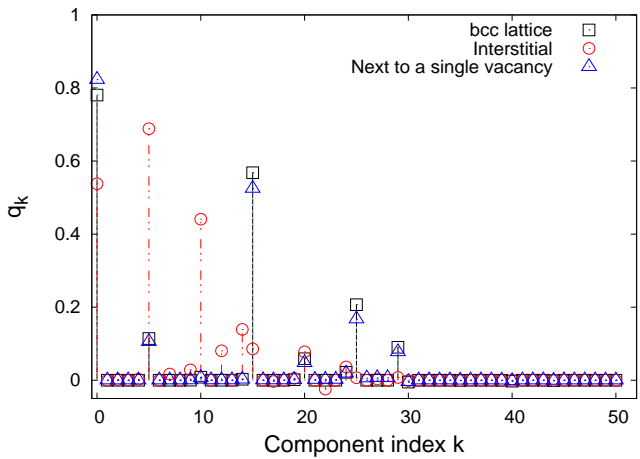}
\caption{(Color on-line) Amplitude of descriptor vectors at 
0 K, as a function of their components for different atomic environments.}
\label{fig:fig_2}
\end{figure}

All the descriptor vectors are computed by QUIP with the SOAP descriptors 
package. We chose the following parameters: $l_{\textrm{max}}=4$,
 $n_{\textrm{max}}=4$, and a $r_{ij}$ cutoff of $0.31$ nm, slightly smaller 
than the unit cell of W with its lattice constant of $0.316$ nm at 0K. 
It is worth mentioning that with this choice all 8 nearest neighbours in the 
tungsten lattice are captured, as well as potential interstitials. 
A larger cut-off radius is possible but some preliminary tests indicated an increased 
tendency of the algorithm towards false positives.
%than the lattice constant of the given material, some functions that 
%describe the atomic environment of each atom will be overlapped and our 
%method will be unable to find actual point defects as first nearest neighbors, 
%rather than distortions in the damaged sample as second or third nearest 
%neighbors.}
Reference DVs $\vec{v}\left(T\right)$ that take into account thermal displacement 
are calculated for thermalized bcc samples at 300 and 600 K. 
These vectors are subsequently used to compute the probability for each atom 
to be in an undistorted environment using Eq. \ref{eq:Eq4}. 
Atoms which exhibit a very low probability
are either potential defects (eg. interstitial) or close to a defect (eg. a vacancy).
The computation of the reference DVs takes $\sim$ 7 sec. for this particular
size of the W sample on a desktop computer. 
%In Fig. \ref{fig:fig_2}, we show the amplitude of the reference DVs as a function 
%of their components.
In figure \ref{fig:fig_2} we show the reference DVs of three standard point defects, 
at 0 K, as a function of their components.
A lattice W atom (without thermal fluctuations), a tungsten 
atom at an interstitial site and an atom next to a single vacancy.
% We notice that the ones for a perfect BCC and for a void/vacancy are similar, 
%with a difference in the magnitude of the nonzero components of the BCC-lattice DV.
%However, the DV for a tetrahedral interstitial atom has more components
%that are not zero due to the different neighborhood of this W atom.
It can be seen that the structure of the non-zero components of the DVs for
a tungsten atom in an ideal environment differs significantly from the structure 
of the DV of a W atom in a tetrahedral interstitial position.
In contrast, the intensity pattern is close for atoms next to a vacancy compared 
to the DV of atoms in an undistorted lattice. 

\begin{figure}[!t]
\centering\includegraphics[width=0.48\textwidth]{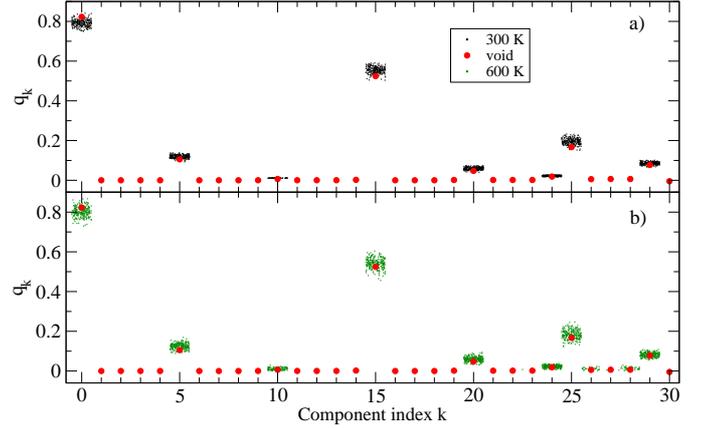}
\caption{ (Color on-line) Amplitude of descriptor vectors 
of a defect free and thermalized W lattices at 300 and 600 K are presented 
in b) and c), respectively. 
The reference DV of a W atom next to a void is included to 
show its difference to the thermalized DVs.
These vector are used as reference descriptors to analyze 
the samples.}
\label{fig:fig_3_new}
\end{figure}

In Fig. \ref{fig:fig_3_new}a) and b) we present the DVs 
of all atom of the defect-free and thermalized BCC samples at 300 and 600 K, 
respectively.
To provide a better visualization of the thermal descriptor vectors, 
we have used a bee-swarm plot: for component indices with intensities ($> 0$), 
the position is given by x$+ 0.5x'$, where $x'$ is a random number 
uniformly distributed in the interval $[-1,1]$. 
It is observed that the distance in vector space at 300 K still allows a good 
identification, because regular lattice atoms which are close in one component 
may differ in another component, Fig. \ref{fig:fig_3_new}a).
The situation is different for a sample at 600 K, as it is shown in 
Fig. \ref{fig:fig_3_new}b). 
Here the overlap is too strong to allow for a reliable identification of atoms 
next to a single vacancy because of DV method is centered around atoms. 
Thus, the DVs for the W sample at 600 K have - as to be expected - a bigger 
variance than those at 300 K.
Nevertheless, the mean of the components of the DVs at 300 and 600 K is very close
to the components of the lattice DV at 0K.
It is therefore less suited for the identification of larger voids (because 
there are no atoms). 
In practice these kinds of defects are more easily detected using Voronoi-based 
methods, e.g. by computing the largest empty sphere around a given location.
In our work, we calculate the number of vacancies in the damaged W sample 
by identifying their spatial location by a k-d-tree algorithm, explained 
in Appendix A.
Thus, a manual inspection of the atoms next to an 
identified single vacancy is recommended to visualize all W atoms around these
kinds of defects, which helps to avoid double-counting of defects (e.g. 8 tungsten atoms 
next to a single vacancy are in a distorted lattice position - but this should be 
counted as 1 defect site only - not as 8).
The lattice W atom reference descriptor vector, the interstitial and the 'atom next to a 
vacancy'-reference descriptor vectors as 
well as their geometry are provided in the supplementary material (Appendix B).
%The probability of finding a W atom in a BCC position is useful to classify 
%and quantify material defects because of allows.
%This analysis guarantees that we are classifying the correct 
%number of defects of the sample. 

%Notice that the vectors' components have different magnitude due to 
%thermal motion. 
%To provide a better visualization of the thermal descriptor vectors, 
%the component index that is different than null is expressed as 
%x$\pm 0.1x'$, where $x'$ is a random number uniformly distributed 
%in the interval $[-1,1]$. 

%We add a threshold of 0.1 on the x axis to the components different than null.

\begin{figure}[!t]
\centering
\includegraphics[trim={0 0 0 0}, clip, width=0.48\textwidth]{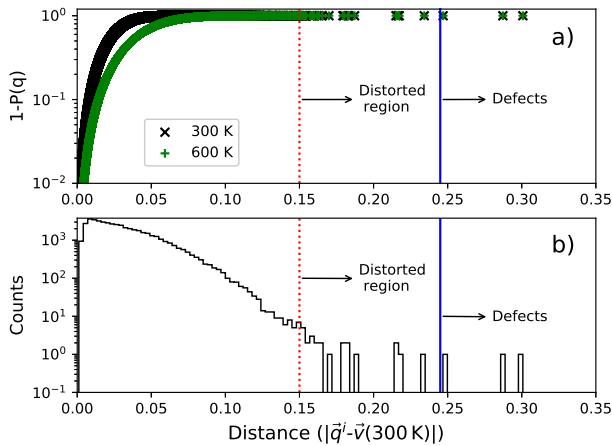}
\caption{(Color on-line) Probability of an atom to be
identified as a defect of the W sample, (Fig. \ref{fig:fig_3}a)
as a function of the distance
difference (Fig. \ref{fig:fig_3}a).
Histogram of the results for the distance difference between
the DV of the mean of the reference W lattice thermalized to 300K, with those for the 
W sample after 1keV collision cascade (Fig. \ref{fig:fig_3}b), on 
the $\langle 0 0 1\rangle$ velocity direction.}
\label{fig:fig_3}
\end{figure}

In Fig. \ref{fig:fig_3}a), we show the probability of an atom to be considered 
at a distorted site, i.e. $1-P(\vec{q})$, as a function of the distance difference by 
using the DVs for a thermalized W sample to 300 ($x$-symbol) and 600K 
($+$-symbol) as a reference.
Already the histogram of the distance from the DV for an atom of 
a bcc-lattice position of a thermalized tungsten lattice provides a good 
hint for a suitable distance value for the defect threshold. 
Here, a manual inspection of the local atom environment for 
different threshold values provides a visualization of the local disorder 
around the atom with the highest probability to be a point defect, which can 
be done by using an open visualization tool like OVITO \cite{Stukowski2009} for example.
Therefore, we noticed that a threshold set at 0.24 gives the number of atoms 
with the highest probability, which are frequently associated with 
the total number of Frenkel pairs. 
However, a threshold set at 0.15 shows us the distorted regions 
around the point defects.

%%%%%%%%%%%%%%%%%%%%%%%%
%%%%%%%%%%%%%%%%%%%%%%%%

%We plot $1-P(q)$ to compare with the results of the 
%Fig. \ref{fig:fig_4}b), which shows the actual defected in the W sample.

%This graph is used to gurantee that the W atoms are well classified as defects 
%of the W sample.
%\textcolor{blue}{ The weighted difference between the amplitudes of the components can 
%computed as:
%\begin{equation}
%r_k^i = \frac{\left| q_k^i -v_k(T) \right|}{\Sigma_{kk}(T)},
%\end{equation}}
%where $q_{k}^{i}$ is the k-th component of the descriptor vector of the $i$th-atom; 
%\textcolor{blue}{ $v_{k}$(T)} is the \textcolor{blue}{mean} descriptor vector 
%for a thermalized sample, this value is defined as the average of the 
%amplitudes of each vector component; 
%\textcolor{blue}{$k$ is used as a component index, as usual; and $\Sigma_{kk}$(T)} 
%is the  amplitude standard deviation of the DV components.
%Then, \textcolor{blue}{as explained in the Sec. 2.3.},
%the distance difference, $d^{\ i}$, is obtained as
%\begin{equation}
%d^{\ i} = \left( \sum_k^N r_k^i \cdot r_k^i \right)^{\frac{1}{2}}.
%\label{eq:distance}
%\end{equation}

In Fig. \ref{fig:fig_3}b) we 
present the results for the euclidean distance between the mean reference descriptor vector $\vec{v}\left(300\,\mathrm{K}\right)$ 
for the thermalized sample at 300 K with the DVs of each atom in the damaged 
pristine W sample after interaction with the 1 keV PKA.
W atoms that have a distance larger than the threshold set to $0.15$ are classified
as atoms being in a distorted environment. 
The subsequent identification of these atoms (e.g. as being close to a vacancy site 
or being an interstitial atom) is done by calculating the 
distance difference between the atomic descriptor vector and the corresponding reference DVs of
the different types of defects.
%W atom next to a vacancy and its interstitial site, respectively.

%%%%%%%%%%%%%%%%%%%%%%%%%%%%%%%%%%%%%%%%%%%%%%%%%%%%%%%%%%%%%%%%%%%%%%
%%%%%%%%%%%%%%%%%%%%%%%%%%%%%%%%%%%%%%%%%%%%%%%%%%%%%%%%%%%%%%%%%%%%%%%
\subsection{Defect classification in a pristine W sample}
For the classification of the type of the distorted environment 
the descriptor vectors of the previously identified atoms are compared with reference
descriptor vectors of a number of common defect structures (like an interstitial atom).
If the distance of the atom descriptor vector and the reference vector of the defect
is \textit{below} some threshold, the atom is correspondingly labelled. 
%We classify (and quantify the number of W atoms as interstitials, a W 
%atom next to a vacancy, or other material defects by 
%computing the distance difference between the DVs of the sample to 
%each reference DV using Eq. \ref{eq:maha}.

\begin{figure}[!t]
\centering
\includegraphics[trim={0 0 0 0},clip,width=0.40\textwidth]{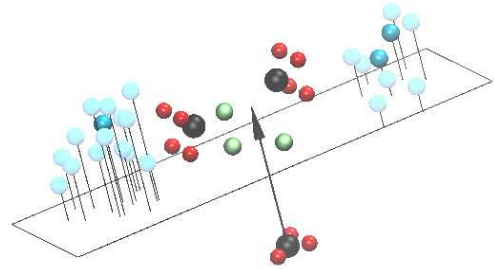}
\caption{(Color on-line) Defects of a pristine W sample after impact of a 
1 keV PKA with a velocity direction of $\langle 0 0 1 \rangle$. 
The W atoms that are identified to be in an distorted environment are resembled 
by spheres using the following color code:
atoms at or close to interstitial sites are indicated in light-blue; 
atoms with the highest probability to be a defects as shown in blue.
; W atoms next to a vacancy are given in red;
atoms being part of a type-a defect are displayed in green; and the 
vacancies as presented in black. 
The arrow indicates the velocity direction of the projectile,
and a plane was introduced for a better perspective view to the figure. 
This image was created by using he Visual Molecular Dynamics (VMD) 
tools \cite{HUMP96}. }
\label{fig:fig_4}
\end{figure}

In Fig. \ref{fig:fig_4}, we show the point defects of the pristine W sample, 
for a PKA velocity direction of $\langle 0 0 1 \rangle$. W atoms that 
are in a defect-free lattice environment have been removed. 
Interstitial atoms in the W sample are represented by blue spheres, and 
the atoms in their distorted local region are shown in light-blue sphere.
The formation of a dumbbell defect, where two atoms share a lattice site 
\cite{PhysRevMaterials.3.043606}, is observed at the right hand side of the 
picture and can be easily identified by the DV for an interstitial site.
W atoms next to a vacancy and type-a defects atoms are presented
as red and green spheres, respectively. 
%Closer inspection reveals that there are three interstitial tungsten atoms 
%(corresponding to the largest distance values in figure \ref{fig:fig_3}b ), 
%each being accompanied by W atoms in the near neighbourhood which are displaced 
%from their lattice position. 
All these atoms are identified by our DV based method.
%by the additional interstitial. Out of the Only three W atoms are considered as potential 
%point defects in the sample, and they are classified as interstitials sites, 
%where the W atoms in their vicinity are clustered on the path of 
%the projectile.
In addition vacancy sites, black spheres, are formed along the trajectory path 
of the primary projectile, and are identified by our method.
%The unclassified atoms are formed around more than a single vacancy, 
%which our DV is unable to detect. 
A black arrow is added to indicate the $\langle 0 0 1 \rangle$ velocity 
direction. This figure was created by using the Visual Molecular Dynamics (VMD) 
tools \cite{HUMP96}.

%%%%%%%%%%%%%%%%%%%%%%%%

\begin{figure}[!b]
\centering
\includegraphics[width=0.35\textwidth]{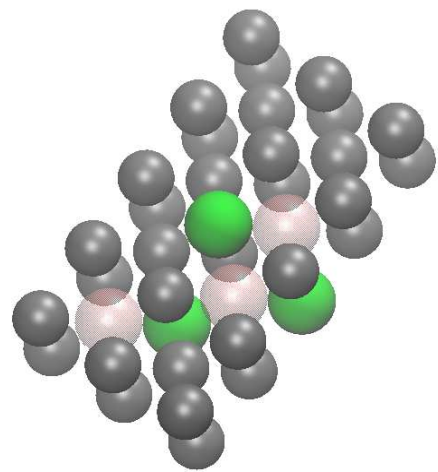}
\caption{(Color on-line) Atomic arrangement of the type-a defect 
(W atoms atom between two vacancies) found by PCA analysis of the 
descriptor vector of a pristine tungsten atom.
 Color code: type-a defect W atoms are represented by green sphere; 
gray and transparent pink spheres shown W lattice atoms and vacancies, respectively.
 }
\label{fig:fig_4_1}
\end{figure}

The approach presented so far does only account for known types of defects, unexpected/unforeseen structures can be also found.
A principle component analysis of the obtained DVs in the various bombardment simulations
(see sec. 3.2 for more details) revealed that besides interstitials and single vacancies 
a third type of a local defect structure was consistently present, which was subsequently labelled 
as type-a defect. 
In the following, the structure of this type-a defect was elucidated by manual 
inspection of the involved atoms and verified by computing again 
the DV of the idealized structure.
In Fig. \ref{fig:fig_4_1} eight bcc unit cells are used to illustrate 
the atomic arrangement of the type-a W defect (green spheres), as a W 
atom between two vacancies (vacancies are represented by transparent pink 
spheres). We note that the W atom can be found either between two consecutive 
vacancies or in a split vacancy.
It is worth mentioning that the DV of a type-a W atom can be 
stored as a new defect reference vector and used in further analysis.
In addition, further formation energy calculations 
can be used to investigate whether or not this kind of defect is indeed stable
or perhaps an unexpected artifact of the MD potential used.

%%%%%%%%%%%%%%%%%%%%%%%%

In order to relate the identified atoms which are in a 
distorted environment (in our case on average 35 W atoms at 
a threshold of $0.15$) to specific classes of defects like Frenkel pairs, 
a subsequent analysis step is necessary. 
In the ideal case (sample temperature at 0K) the 8 W atoms which surround a single 
vacancy identified by our method are all counted as being in a single empty lattice 
site environment. 
However, our method identifies $2-5$ W atoms around a vacancy due to thermal motion.
Thus, by applying this post-processing to the case of the pristine W sample, 
we obtain an average number of 3 Frenkel pairs and the correct visualization 
of the W atoms around the vacancies.
In table \ref{tab:tab1}, we list the total number of W atoms that are
identified as being located at imperfect sites; the percentages of interstitial sites,
W atoms next to a vacancy, and W atoms between two vacancies (type-a) for
different velocity directions at a threshold of $0.15$, while the number of atoms at the highest probability are reported in parentheses at a threshold of $0.24$.
In the same table, we report the average of the obtained results for the 
seven MD simulations, that are performed for random velocity directions.
The total number of defects for each velocity direction are: 
$\{ 36 \ (3), 34 \ (3), 36 \ (3),33 \ (3),35 \ (3),31 \ (2),33 \ (2)\}$. 
The majority of the sample defects, found by our method, are interstitial sites 
with several atoms in their local atomic environment. A minority of the defects
is consistently appearing as type-a.
A comparison with the results of A. E. Sand et al. 
\cite{SAND2016119} and K. Nordlund et al. \cite{Kai_N} who obtained 3 Frenkel pairs shows  
good agreement at 1 keV of PKA. 
The result can be also compared to the expected number of Frenkel pairs (FP)
as given Setyawan et al. \cite{SETYAWAN2015329} : 
\#FP = $0.49 \left(E_{\textrm{PKA}}/E_{\textrm{d}}\right)^{0.74} = 2.24$
at a sample temperature of 300 K with $E_{\textrm{d}} = 128$ eV.
 In the supplementary material, we provide a video to visualize the 
sample defects under different angles.
%Nevertheless, the authors used the Wigner-Seitz method to compute 
%the number of Frenkel pairs, where defects like the type-a defect are difficult to 
%find and demonstrate an advantage of our DV based method to identify 
%material defects.

\begin{table}[!t]
\centering
\begin{tabular}{l c c c c}
\hline \hline
\textbf{Vel. dir.} & \textbf{Inter.}  & \textbf{Next to vac. (Vac.)} &
\textbf{Type-a}  & \textbf{Total}\\
\hline
$\langle 0 0 1 \rangle$     & 24 (3) & 7 (3)   & 2 & 33 \\
$\langle 0 1 1 \rangle$     & 28 (2)  & 6 (2)  & 2 & 36 \\
$\langle 1 1 1 \rangle$     & 24 (2)  & 7 (2)  & 4 & 35 \\
%$\langle \frac{1}{2} 0 \frac{1}{2} \rangle$     & 75.3  & 19.4  & 5.3 & 36 \\
$\langle r_1 r_2 r_3 \rangle$     & 25 (3)  & 7 (3)  & 2 & 34 \\
\hline
Average                     & 25 (3)  & 7 (3)  & 3 & 35 \\
\hline \hline
\end{tabular}
\caption{Defect quantification of pristine W samples, after 
collision cascade, at different velocity directions.
$r_i$ with $i=1,2,3$ are uniform random numbers
in the interval $[0,1]$. 
The number of atoms at the highest probability are reported in parentheses.}
\label{tab:tab1}
\end{table}

%%%%%%%%%%%%%%%%%%%%%%%%%%%%%%%%%%%%%%%%%%%%%%%%%%%%%%%%%%%%%%%%%
%%%%%%%%%%%%%%%%%%%%%%%%%%%%%%%%%%%%%%%%%%%%%%%%%%%%%%%%%%%%%%55
\subsection{Defect classification of deuterated W samples  }
\label{sec:deuterium_defects}

\begin{figure}[!b]
\centering
\includegraphics[width=0.48\textwidth]{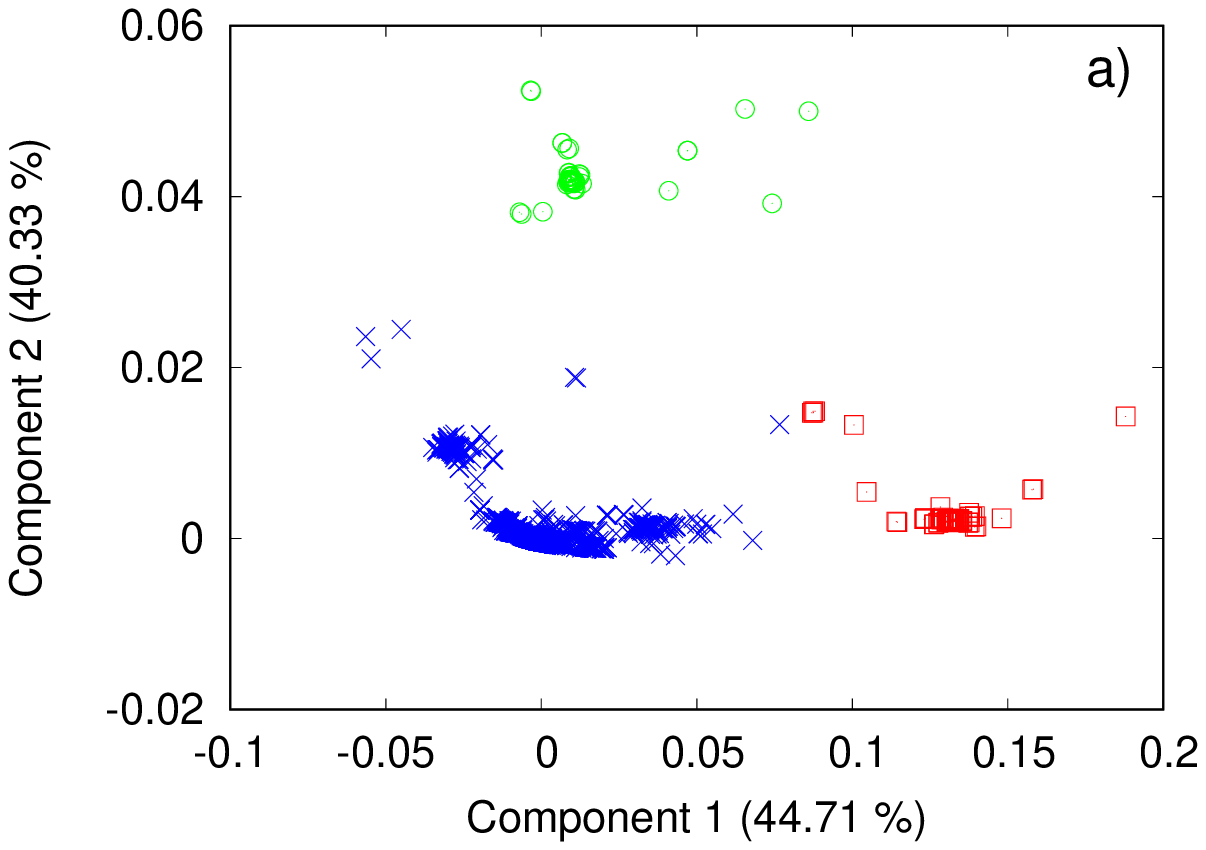}
\includegraphics[width=0.48\textwidth]{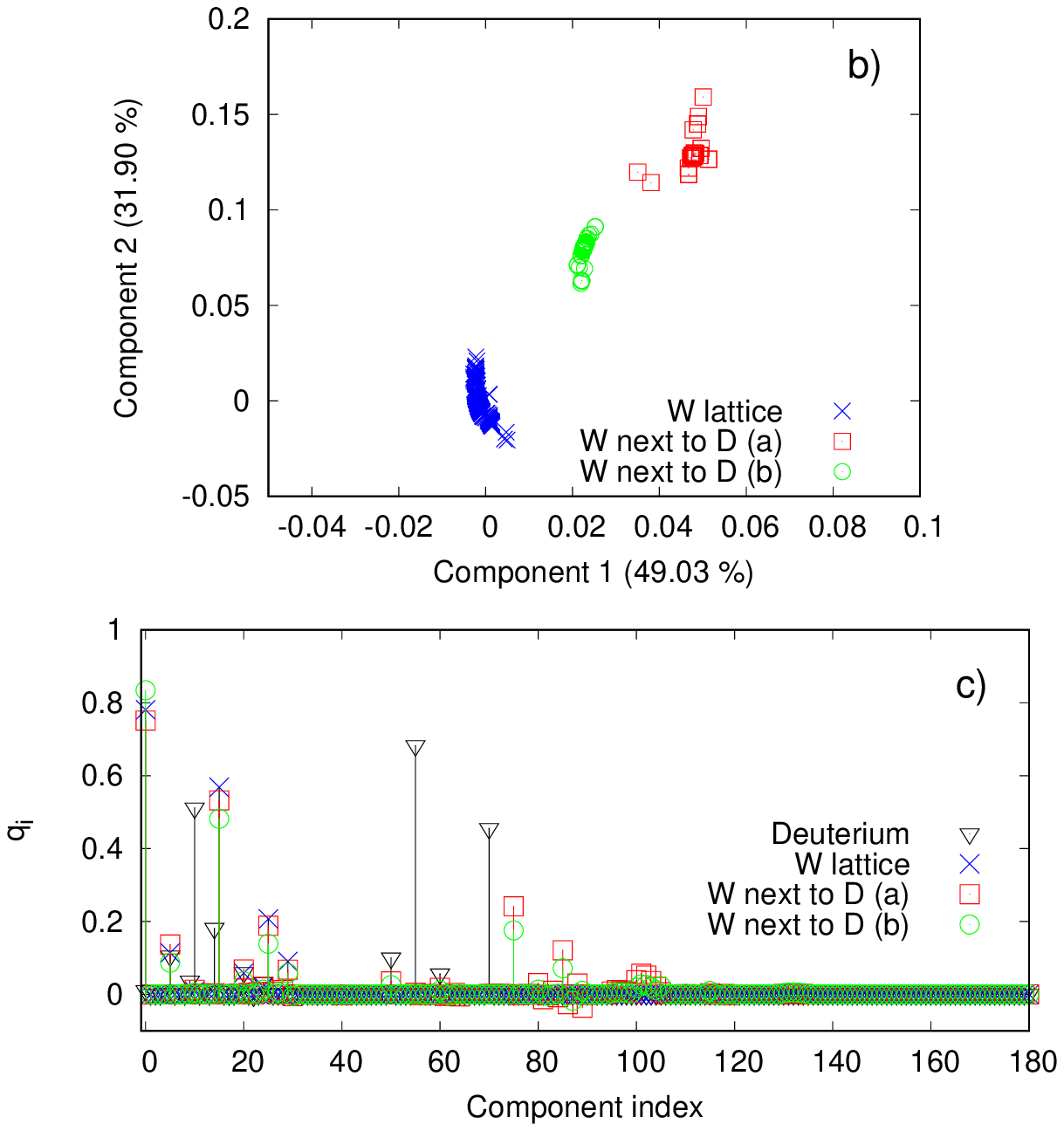}
\caption{(Color on-line) PCA results and DVs of the deuterated 
sample for the first approach in a) and for the second approach in b).
The DVs are shown as a function of the component index for the second 
approach in c).}
\label{fig:fig_5}
\end{figure}

The defect analysis of multi-component samples (here with W and D)
can in principle follow two different approaches. 
In the first approach one tries to get along with the DVs for 
tungsten only, indirectly accounting for the displacement due 
to presence of interstitial D atoms. 
The displacement has to be taken into account because it happens 
also in defect-free systems.
Initially, the reference DVs for the tungsten atoms surrounding 
an interstitial deuterium atom are computed by keeping the W 
atoms positions fixed after (virtually) removing the interstitial 
D atoms. 
This allows to remain the computation of the DV identical to 
the mono-species case. 
However, this approach will fail for larger interstitial atoms 
because the atom displacement becomes too large, and will be
cumbersome for systems with more than two species.
We identify the DVs for a W lattice atom and a W atom next 
to a D using a principal component analysis (PCA).
As shown in Fig. \ref{fig:fig_5}a) all the projected DV (referring 
to defect-free tungsten with interstitial D atom) show a 
considerable scatter. The reason for this is simple: the 
interstitial deuterium displaces the W atoms from their 
equilibrium bcc-lattice positions, which complicates the 
subsequent identification of defects and can result in 
false classifications.
The quantification and classification of the point defects,
found in the damaged sample by this first approach, are reported
in the table \ref{tab:tab2}.
Please note that a number of W atoms next to a vacancy turned out
to be incorrectly labelled by this first approach, which we 
therefore do not recommend - despite its simplicity.
%Besides that, the visualization of this point defects are 
%presented in the supplementary material, where the PKA 
%velocity direction is $\langle 0 0 1 \rangle$ and it is 
%observed that several W atoms are next to 
%a D atom, wrongly identified as W atom next to a vacancy.

\begin{table}[!b]
\centering
\begin{tabular}{l c c c c}
\hline
\textbf{Vel. dir.} & \textbf{Inter.}  & \textbf{Vac.} &
\textbf{New type} & \textbf{Total}\\
\hline
\multicolumn{5}{c}{Approach 1} \\
\hline
$\langle 0 0 1 \rangle$     & 3 (1) & 5 (1) & 35 & 43 \\
$\langle 0 1 1 \rangle$     & 3 (1) & 5 (1) & 36 & 44 \\
$\langle 1 1 1 \rangle$     & 3 (1) & 5 (1) & 33 & 41   \\
%$\langle \frac{1}{2} 0 \frac{1}{2} \rangle$     & 6.5 & 10.9 & 82.6  & 46  \\
$\langle r_1 r_2 r_3 \rangle$     & 3 (1) & 5 (1)  & 37  &  45  \\
Average                     & 3 (1) & 5 (1)  & 35  &  43  \\
\hline
\multicolumn{5}{c}{Approach 2} \\
\hline
$\langle 0 0 1 \rangle$     & 3 (1) & 1 (1) & 31 & 35 \\
$\langle 0 1 1 \rangle$     & 3 (1) & 1 (1) & 33 &  37 \\
$\langle 1 1 1 \rangle$     & 3 (1) & 1 (1) & 30 & 34   \\
%$\langle \frac{1}{2} 0 \frac{1}{2} \rangle$     & 7.7 & 2.6 & 89.7  &  39  \\
$\langle r_1 r_2 r_3 \rangle$     & 3 (1) & 1 (1)  & 31  &  35  \\
Average                     & 3 (1) & 1 (1)  & 31  &  35  \\
\hline
\end{tabular}
\caption{Defect quantification of a deuterated damaged W samples
for different PKA velocity directions. 
Total number of Frenkel pairs are reported in parentheses.}
\label{tab:tab2}
\end{table}

%In the following section, we present the analysis of the defects 
%identification of a deuterated sample needs a computation of  W atoms that
% are around to a D in advance, otherwise they can be miscounted as defects 
%in the sample.
%We considered the following cases:

%\begin{enumerate}
%\item All the D atoms are removed from the sample to compute the 
%reference DVs for a W lattice and a W atom next to a deuterium.
%This allows us to follow the previous analysis. 
%\item The D atoms are included in the computation of the DVs, 
%and further analysis has to be done.
%\end{enumerate}

Since the first approach does not provide viable results,
the information about the D atoms needs to be included in 
the computation of new multi-component DVs for W atoms next 
to a D atom. 
This second approach is straightforward but increases the number
of basis functions and the nature of the reference DVs.
In Fig. \ref{fig:fig_5}b), we show the PCA results of the DVs 
of the damaged-free and deuterated W sample, where the first 
and second principal components capture $49.03$ \% and $31.90$
\% of the variance of the data.
%Representing the 81 \% of the information of the descriptor vectors.
We notice that the cluster around the origin is related to 
the DVs of a lattice W atom in a defect-free and pristine
environment. 
There are two clearly separated clusters with centers at the
points $(0.021, 0.08)$ and $(0.055,0.13)$, that are associated 
to W atoms next to the D atom as a first and second neighbors,
respectively. 
Closer inspection reveals that W atoms that are next to 
an octahedral interstitial D atom can be assigned to a
first-neighbor DV, and W atoms next to a tetrahedral D atom 
are identified by a second-neighbor DV.

In Fig. \ref{fig:fig_5}c), we show the DVs for a lattice W 
atom ($\times$ symbol); for a W atom next to a D atom, 
where (a) are first neighbors ($\square$) and (b) are 
considered as second neighbors ($\circ$); and for a D atom
($\triangledown$). 
Due to the inclusion of the D atom in the DVs computation, 
more components are needed to describe the D atom in the local
neighborhood of the W atoms resulting in DVs with $k=181$
$(0\ldots 180)\,$ components.
To be consistent with the number of components, 
the reference DVs for the pristine W case are extended 
to 181 components by adding additional vector components with
their value set to zero.
%It is worth noticing that the first components of the DV of a W atoms
% are similar in both cases, the effect of the D atoms is visible in the 
%components 70-110.
%The former classification could be considered as second neighbors in the 
%vicinity of the D atoms.
We then used these new DVs to identify W atoms next to D atoms in 
deuterated W samples. An example for that is given in Fig. 
\ref{fig:fig_1}.

%%%%%%%%%%%%%%%%%%%%%%%%
\begin{figure}[!t]
\centering
\includegraphics[width=0.48\textwidth]{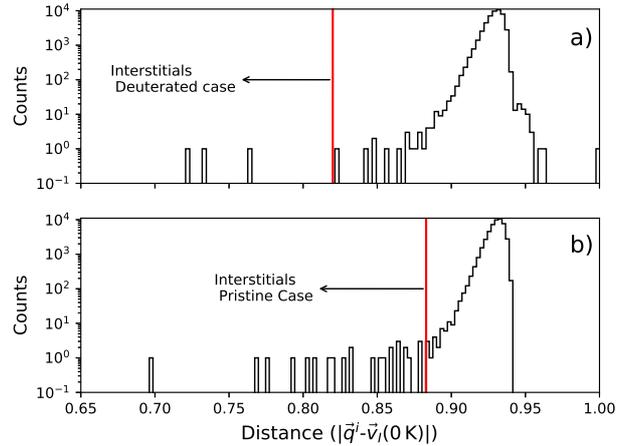}
\caption{(Color on-line) Histogram of the distance 
between the atom DVs and the reference DV $\vec{v}_{I}\left(0\,\mathrm{K}\right)$ for the
\textit{deuterated} case, in a), and for the \textit{pristine} 
case in b).
The velocity direction for the considered sample is 
$\langle 0 0 1 \rangle$ in both cases.
Please note that in both panels only interstitial defects are identified - although the pristine sample has eventually fewer atoms in a distorted environment, the number of Frenkel pairs is larger.}
\label{fig:fig_hist_6}
\end{figure}

To analyze the deuterated and damaged W sample, we first 
identified the W atoms that are in interstitial sites. 
Following the approach described in the previous section, 
a histogram is computed for the distances of the atoms DVs to the 
reference DV for an \textit{interstitial} atom
$\vec{v}_{I}\left(0\,\mathrm{K}\right)$, and shown in
Fig. \ref{fig:fig_hist_6}a).
A comparison of the same quantity for the pristine case is
given in Fig. \ref{fig:fig_hist_6}b). 
The number of interstitial atoms or atoms next to an interstitial is larger for the pristine case. However, the picture changes if all atoms in a distorted environment are considered. As is detailed below, the additional deuterium in the sample changes the relative occurrence of the observed defect-types.  
%the histogram forof the distance 
%$difference between the chosen reference DV to the deuterated sample in a), 
%results for a pristine sample are displayed in b) to show the effect of the 
%D atom in the formation of interstitials. 
%Note that both graphs show the same number of instertitials in the sample 
%with the same threshold.

We tabulated the number of defects in the deuterated W sample as 
a function of the PKA velocity direction in the Tab. \ref{tab:tab2}, 
where total number of Frenkel pair is reported in parentheses and 
the total number of defects for the random velocity directions 
are $\{ 35 (1),36 (1),36 (1),35 (1),31 (1),35 (1)\}$.
Comparing the results of Sec. 3.1 (pristine W) with the present data 
for a sample with D, the results reveal a small but persistent 
difference in the total number of atoms in a distorted environment 
between the two samples.

%Also, it is observed that some D atoms do not affect the sample due to W atoms
%are not in its vicinity.
We provide a movie of the classification of defects 
for a deuterated and damaged W sample at a $\langle 0 0 1 \rangle$ velocity direction in the supplementary material. 
A new type of defect can be identified for the deuterated case after the collision cascade: 
Almost all ($> 80\%$)
of the displaced tungsten atoms are accompanied by a deuterium in the near vicinity (c.f. table \ref{tab:tab2}, approach 2).
%, for the deuterated case in 
%the second approach, is observed as a non-lattice W atom 
%with a deuterium atom in its vicinity.
%A 2D snapshot of the identified point defects, with the presence of 
%D atoms in the sample, does not show enough information about the classification 
%of the defects and a 3D visualization is needed, which is presented in 
%the supplementary material.
In addition, for all simulations except one the collision 
cascades yield a larger number of defects in the deuterated 
sample compared to the pristine one. 
Although the statistics is not very good, such a results 
should happen by chance in less than 1 \% of all cases if the 
defect probability is the same for both cases.

We conclude that the effect of the presence of deuterium in the 
W sample is modest but that on average the number of point defects
defects is larger than in the pristine one. 
This could point towards a stabilization of defects by hydrogen 
- but here many more simulations are needed to substantiate 
this hypothesis.
%The material defects that are found for the second case are presented in 
%Further analysis has to be done to identify sample defects with the D atom 
%included due to their geometry, which is beyond of the scope of this paper.

\section{Concluding Remarks}
\label{conclusions}
MD simulations are commonly used to study radiation 
damage in crystalline materials. 
The analysis of the results are sometimes misleading by the 
relative small volume fraction of the modified parts of the 
sample and by the formation of unforeseen defect types after
irradiation. 
In this work, we present a fingerprint method capable to identify
defects in crystalline samples. It is based on a descriptor 
vector of the local environment of an atom. 
The DV is sensitive to local modifications of the atomic 
configuration but is insensitive to global changes (like 
rotations) at the same time.
The proposed approach is suited to identify and classify defects 
of a given sample in a semi-automated manner. 
This opens the door for a semi-automated continuous monitoring 
of the defect evolution during collision cascades - something 
which has rarely been done so far because of the large manual 
effort involved.

It provides also a probabilistic quantification for each atom 
of the sample to be in a distorted local environment.
We applied our method to irradiated pristine and deuterated 
W samples. 
For this, we used QUIP to compute the descriptor vectors of all
atoms of the samples to describe the neighborhood of each atom. 
The MD simulations have been performed 
with LAMMPS using a primary tungsten knock-on atom  of 1 keV 
and sample temperature of 300 K.
%We also provide the descriptor vector for some of the most common cases like 
%an atom in its body centered cubic location, a tungsten atom at an interstitial 
%site, and an atom next to a \textcolor{blue}{single} vacancy.  
For the deuterated case, we used a principal component analysis 
to identify suitable descriptor vectors for W atom as first and 
second nearest neighbors of D atoms. 
%We also found two atom configurations for a W atom next to a D. 
These DVs are subsequently used to identify and quantify 
defects in the  D-W samples after irradiation.
It turns out that the deuterated samples exhibit, with high
probability, more defects than the pristine samples - although
improved statistics is needed to substantiate this claim further. 
Future work is on the one hand side focused on the effect of the
impact energy and the deuterium concentration on the formation 
of defects in tungsten.
The other line of research addresses the identification of new 
types of defects and the application of the DV based approach 
to other systems.

\section*{Acknowledgments}
F.J.D.G gratefully acknowledges funding from A. von Humboldt 
Foundation and C. F. von Siemens Foundation for research fellowship.
We would also like to acknowledge the input of the anonymous 
reviewers which has improved the quality of the paper.
Simulations were performed using the LINUX cluster 
at the Max-Planck Institute for plasma physics.

\section*{Appendix A. Vacancies identification}
\begin{figure}[!t]
\centering
\includegraphics[width=0.48\textwidth]{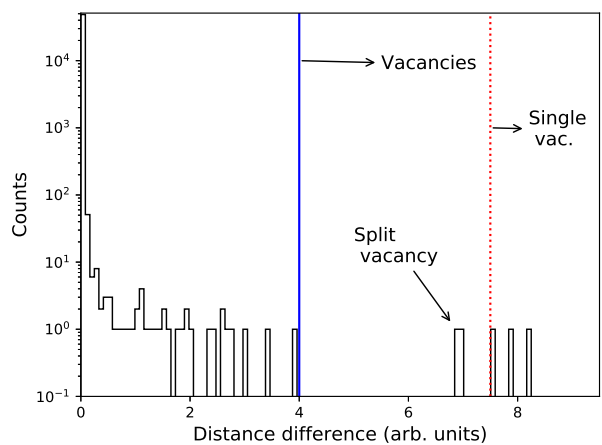}
\caption{(Color on-line) Histogram of distance difference 
between atom positions of a damaged sample and a thermalized one. 
Single vacancies and a split-vacancy are clearly identified by our method}
\label{fig:A1_hist}
\end{figure}

The number of vacancies is calculated by
computing the nearest neighbor distance to the closest atom 
of a thermalized defect-free W sample and the damaged sample, 
by using the k-d-tree algorithm
\cite{Bentley:1975:MBS:361002.361007}.  
A comparison between  atom-atom  distances  indicates  the 
vacancy location in  the damaged  sample,  where distance 
values  exceed  a  lattice  dependent  threshold. 
In Fig. \ref{fig:A1_hist} we present a histogram of the results 
of identification of vacancies for a velocity direction of 
$\langle 0 0 1 \rangle$. 
We notice that the distance difference values beyond a threshold 
set at 4 are associated to the location of vacancies.
In order to obtain a visualization of all the defects
classification, the W atoms that are next to these vacancies 
are identified and shown in the final analysis, Fig. \ref{fig:fig_4}.
%%%%%%%%%%%%%%%%%%%%%%%%

\section*{Appendix B. Supplementary material}
\label{sup_mat:B}
\href{https://data.mendeley.com/datasets/3gdffbrzhs/draft?a=9e56a9c8-aced-4977-844e-9a755bbfb060}{DVs and visualization videos}

\section*{References}

\bibliography{bibliography}

\begin{thebibliography}{10}
\expandafter\ifx\csname url\endcsname\relax
  \def\url#1{\texttt{#1}}\fi
\expandafter\ifx\csname urlprefix\endcsname\relax\def\urlprefix{URL }\fi
\expandafter\ifx\csname href\endcsname\relax
  \def\href#1#2{#2} \def\path#1{#1}\fi

\bibitem{doi:10.13182/FST11-A12443}
J.~Alvarez, A.~Rivera, R.~Gonzalez-Arrabal, D.~Garoz, E.~del Rio, J.~M.
  Perlado, Materials research for hiper laser fusion facilities: Chamber wall,
  structural material and final optics, Fusion Science and Technology 60~(2)
  (2011) 565--569.

\bibitem{BOLT200243}
H.~Bolt, V.~Barabash, G.~Federici, J.~Linke, A.~Loarte, J.~Roth, K.~Sato,
  Plasma facing and high heat flux materials â€“ needs for {I}{T}{E}{R}
  and beyond, Journal of Nuclear Materials 307-311 (2002) 43 -- 52.

\bibitem{mayer}
M.~Mayer, M.~Andrzejczuk, R.~Dux, E.~Fortuna-Zalesna, A.~Hakola, S.~Koivuranta,
  K.~Krieger, K.~J. Kurzydlowski, J.~Likonen, G.~Matern, Tungsten erosion and
  redeposition in the all-tungsten divertor of {ASDEX} upgrade, Physica Scripta
  T138 (2009) 014039.

\bibitem{Atsuyuki}
A.~Okabe, B.~Boots, K.~Sugihara, S.~N. Chiu, Spatial tessellations: concepts
  and applications of Voronoi diagrams, John Wiley and Sons, Inc., New York,
  NY, 2000.

\bibitem{doi:10.1063/1.4849775}
Y.-N. Liu, T.~Ahlgren, L.~Bukonte, K.~Nordlund, X.~Shu, Y.~Yu, X.-C. Li, G.-H.
  Lu, Mechanism of vacancy formation induced by hydrogen in tungsten, AIP
  Advances 3~(12) (2013) 122111.

\bibitem{PhysRevLett.109.095505}
E.~A. Lazar, J.~K. Mason, R.~D. MacPherson, D.~J. Srolovitz, Complete topology
  of cells, grains, and bubbles in three-dimensional microstructures, Phys.
  Rev. Lett. 109 (2012) 095505.

\bibitem{IAEA}
I{A}{E}{A} {C}hallenge on materials for fusion,
  \url{https://challenge.iaea.org/challenges/2018-NA-Mat-Fusion}, accessed:
  March 18th, 2019.

\bibitem{PhysRevB.87.184115}
A.~P. Bart\'ok, R.~Kondor, G.~Cs\'anyi, On representing chemical environments,
  Phys. Rev. B 87 (2013) 184115.

\bibitem{PhysRevB.90.104108}
W.~J. Szlachta, A.~P. Bart\'ok, G.~Cs\'anyi, Accuracy and transferability of
  gaussian approximation potential models for tungsten, Phys. Rev. B 90 (2014)
  104108.

\bibitem{quip}
\url{http://libatoms.github.io/QUIP/} (2018).

\bibitem{KAUFMANN2007521}
M.~Kaufmann, R.~Neu, Tungsten as first wall material in fusion devices, Fusion
  Engineering and Design 82~(5) (2007) 521 -- 527.

\bibitem{PIAGGI2015233}
P.~Piaggi, E.~Bringa, R.~Pasianot, N.~Gordillo, M.~Panizo-Laiz, J.~del RÃ­o,
  C.~G. de~Castro, R.~Gonzalez-Arrabal, Hydrogen diffusion and trapping in
  nanocrystalline tungsten, Journal of Nuclear Materials 458 (2015) 233 -- 239.

\bibitem{CHEN2016190}
Z.~Chen, L.~J. Kecskes, K.~Zhu, Q.~Wei, Atomistic simulations of the effect of
  embedded hydrogen and helium on the tensile properties of monocrystalline and
  nanocrystalline tungsten, Journal of Nuclear Materials 481 (2016) 190 -- 200.

\bibitem{Maha}
P.~Mahalanobis, On tests and measures of group divergence {I}. theoretical
  formulae, J. and Proc. Asiat. Soc. of Bengal 26 (1930) 541.

\bibitem{SETYAWAN2015329}
W.~Setyawan, G.~Nandipati, K.~J. Roche, H.~L. Heinisch, B.~D. Wirth, R.~J.
  Kurtz, Displacement cascades and defects annealing in tungsten, part {I}:
  Defect database from molecular dynamics simulations, Journal of Nuclear
  Materials 462 (2015) 329 -- 337.

\bibitem{Wright_nucl_fusion}
G.~M. Wright, M.~Mayer, K.~Ertl, G.~de~Saint-Aubin, J.~Rapp, Hydrogenic
  retention in irradiated tungsten exposed to high-flux plasma, Nucl. Fusion 50
  (2010) 075006.

\bibitem{Herrmann_nucl_fusion}
A.~Herrmann, H.~Greuner, N.~Jaksic, M.~Balder, A.~Kallenbach, et~al., Solid
  tungsten divertor-{I}{I}{I} for {A}{S}{D}{E}{X} {U}pgrade and contributions
  to {I}{T}{E}{R}, Nucl. Fusion 55 (2015) 063015.

\bibitem{PhysRevB.17.1302}
T.~Schneider, E.~Stoll, Molecular-dynamics study of a three-dimensional
  one-component model for distortive phase transitions, Phys. Rev. B 17 (1978)
  1302.

\bibitem{DOMINGUEZGUTIERREZ201756}
F.~Dom\'inguez-Guti\'errez, P.~Krsti\'c, Sputtering of lithiated and oxidated
  carbon surfaces by low-energy deuterium irradiation, Journal of Nuclear
  Materials 492 (2017) 56 -- 61.

\bibitem{PLIMPTON19951}
S.~Plimpton, Fast parallel algorithms for short-range molecular dynamics,
  Journal of Computational Physics 117~(1) (1995) 1 -- 19.

\bibitem{Juslin}
N.~Juslin, P.~Erhart, P.~Tr\"askelin, J.~Nord, K.~O.~E. Henriksson,
  K.~Nordlund, E.~Salonen, K.~Albe, Analytical interatomic potential for
  modeling nonequilibrium processes in the {W}-{C}-{H} system, Journal of
  Applied Physics 98~(12) (2005) 123520.

\bibitem{1402-4896-2011-T145-014036}
U.~von Toussaint, S.~Gori, A.~Manhard, T.~H\"oschen, C.~H\"oschen, Molecular
  dynamics study of grain boundary diffusion of hydrogen in tungsten, Physica
  Scripta 2011~(T145) (2011) 014036.

\bibitem{FU2018278}
B.~Fu, M.~Qiu, J.~Cui, M.~Li, Q.~Hou, The trapping and dissociation process of
  hydrogen in tungsten vacancy: A molecular dynamics study, Journal of Nuclear
  Materials 508 (2018) 278 -- 285.

\bibitem{FU2018}
B.~Fu, M.~Qiu, L.~Zhai, A.~Yang, Q.~Hou, Molecular dynamics studies of
  low-energy atomic hydrogen cumulative bombardment on tungsten surface,
  Nuclear Instruments and Methods in Physics Research Section B: Beam
  Interactions with Materials and Atoms (In press).

\bibitem{Stukowski2009}
A.~Stukowski, Visualization and analysis of atomistic simulation data with
  {OVITO}{\textendash}the open visualization tool, Modelling and Simulation in
  Materials Science and Engineering 18~(1) (2009) 015012.

\bibitem{PhysRevMaterials.3.043606}
P.-W. Ma, S.~L. Dudarev, Symmetry-broken self-interstitial defects in chromium,
  molybdenum, and tungsten, Phys. Rev. Materials 3 (2019) 043606.

\bibitem{HUMP96}
W.~Humphrey, A.~Dalke, K.~Schulten, {VMD} -- {V}isual {M}olecular {D}ynamics,
  Journal of Molecular Graphics 14 (1996) 33--38.

\bibitem{SAND2016119}
A.~Sand, J.~Dequeker, C.~Becquart, C.~Domain, K.~Nordlund, Non-equilibrium
  properties of interatomic potentials in cascade simulations in tungsten,
  Journal of Nuclear Materials 470 (2016) 119.

\bibitem{Kai_N}
K.~Nordlund, S.~J. Zinkle, A.~E. Sand, F.~Granberg, et~al., Improving atomic
  displacement and replacement calculations with physically realistic damage
  models, Nature Communications 9 (2018) 1084.

\bibitem{Bentley:1975:MBS:361002.361007}
J.~L. Bentley, Multidimensional binary search trees used for associative
  searching, Commun. ACM 18~(9) (1975) 509--517.

\end{thebibliography}

\end{document}